\documentclass[twocolumn,prb,showpacs,floatfix]{revtex4}

\usepackage{graphicx}
\usepackage{bm}
\usepackage{color}
\usepackage{amsmath}
\usepackage{amsfonts}
\usepackage{amssymb}
\usepackage{epstopdf}
\usepackage{ulem}

\begin{document}

\preprint{APS/123-QED}

\title{Pressure-induced superconductivity in EuFe$_2$As$_2$ without a quantum critical point: magnetotransport and upper critical field measurements under high pressure}

\author{Nobuyuki Kurita$^{1,2}$\footnote{Present address: Department of Physics, Tokyo Institute of Technology, Meguro-ku, Tokyo 152-8551, Japan}, Motoi Kimata$^{3}$, Kota Kodama$^{1,4}$, Atsushi Harada$^1$, Megumi Tomita$^1$, \\
 Hiroyuki S. Suzuki$^1$, Takehiko Matsumoto$^1$, Keizo Murata$^5$, Shinya Uji$^{1,2,4}$, and Taichi Terashima$^{1,2}$}
\affiliation
{
$^1$National Institute for Materials Science, Tsukuba, Ibaraki 305-0003, Japan \\
$^2$JST, Transformative Research-Project on Iron Pnictides (TRIP), Chiyoda, Tokyo 102-0075, Japan \\
$^3$Institute for Solid State Physics, The University of Tokyo, Kashiwanoha, Kashiwa, Chiba 277-8581, Japan \\
$^4$Graduate School of Pure and Applied Sciences, University of Tsukuba, Ibaraki 305-0003, Japan \\
$^5$Department of Physics, Graduate School of Science, Osaka City University, Osaka 558-8585, Japan
}
\date{\today}

\begin{abstract}

Resistivity and Hall effect measurements of EuFe$_2$As$_2$ up to 3.2\,GPa 
indicate no divergence of quasiparticle effective mass at the pressure $P_\mathrm{c}$ 
where the magnetic and structural transition disappears.
This is corroborated by analysis of the temperature ($T$) dependence of the upper critical field.
$T$-linear resistivity is observed at pressures slightly above $P_\mathrm{c}$.
The scattering rates for both electrons and holes are shown to be approximately $T$-linear.
When a field is applied, a $T^2$ dependence is recovered, 
indicating that the origin of the $T$-linear dependence is spin fluctuations.
\end{abstract}


\maketitle

\section{Introduction}

Since the discovery of superconductivity in LaFeAs(O, F) at $T_\mathrm{c}$\,=\,26\,K,\cite{Kamihara}
considerable attention has been paid to iron-based superconductors (SCs) with a variety of crystal structures 
containing stacked iron-pnictide (or -chalcogenide) layers.\cite{review}
The maximum values of $T_\mathrm{c}$ thus far achieved are 54\,-\,56\,K\,\cite{Kito2008, ZARen2008a,Wang_56K} 
and 39\,K\,\cite{Rotter2008b} in the ``1111" ($R$FeAsO; $R$\,=\,rare earth) 
and ``122" ($A$Fe$_2$As$_{2}$; $A$\,=\,alkaline earth or Eu) groups, respectively.
Despite intensive research, the detailed mechanism of the superconductivity, 
for example, the symmetry of the SC order parameter, remains highly controversial.\cite{Mazin_s+-,Kuroki_s+-,NakaiJPSJ2008,Hashimoto_PRBR2010,Kontani_s++,Sato_s++}
It has been revealed that iron-based SCs have a unique Fermi surface (FS) structure, 
typically consisting of two- or three-hole and two-electron sheets.\cite{Singh2008PRL,Ding2008EPL}
The 1111 and 122 parent compounds undergo FS reconstruction associated with 
an antiferromagnetic (AF) order of Fe moments at $T_\mathrm{0}$.\cite{review}
With the suppression of $T_\mathrm{0}$ via dopings\,\cite{Kamihara,Rotter2008b} or 
the application of pressure ($P$),\cite{Alireza} the superconducting (SC) ground state can be triggered.
Hence, magnetic instability may play an important role in iron-based SCs.

One of the intriguing issues of iron-based SCs is the origin of 
non-Fermi-liquid (NFL) behavior in their transport properties, such as 
$T$-linear resistivity,\cite{Liu_NFL,Gooch_NFL}
which emerges as $T_\mathrm{0}$ is suppressed.
The existence of a quantum critical point (QCP), where the second-order transition temperature becomes zero, 
in iron-based SCs has theoretically been proposed\,\cite{Dai_PNAS2009}
and has been demonstrated by the observation of a peak in the penetration depth, 
which is proportional to $(n/m^*)^{-1/2}$, at the optimal doping in the BaFe$_2$(As,P)$_2$ system,\cite{Hashimoto_Science2012} for example ($n$ and $m^*$ are the carrier number 
and quasiparticle effective mass, respectively).

However, the existence of a QCP does not appear to be universal in iron-based SCs nor is 
its relevance to the superconductivity clear, as suggested by phase diagrams of 
La-based 1111 systems,\cite{Luetkens_NatureM2009,Iimura_Nature2012} or the composition $x$ dependences of the 
penetration depth and Drude weight of optical conductivity, both of which are related to $n/m^*$, 
in the Ba(Fe$_{1-x}$Co$_x$)$_2$As$_2$ system,\cite{Nakajima_PRB2010,Gordon_PRB2010} for example.
In addition, it has been argued that the interpretation of the NFL-like behavior
may not be straightforward owing to the multiband character of iron-based systems.\cite{Fang_PRB2009,Albenque_PRL}

\begin{figure}
\begin{center}
\includegraphics[width=0.9\linewidth]{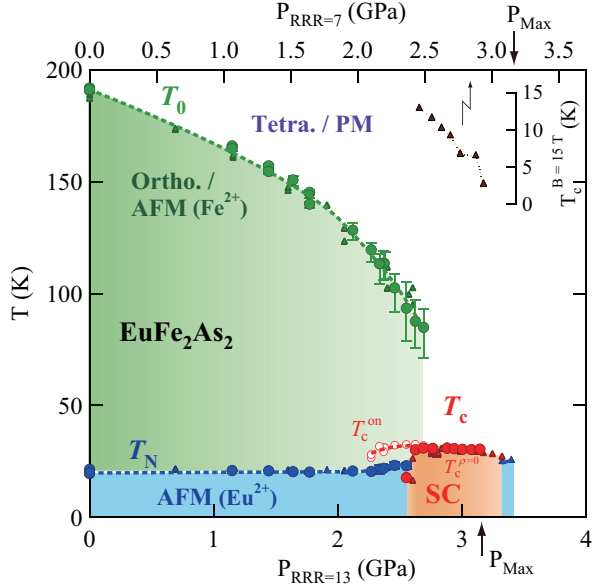}
\end{center}
\caption{(Color online) $P$\,$-$\,$T$ phase diagram in EuFe$_2$As$_2$ with $RRR$\,=\,13 (circles, bottom axis) 
and 7\,\cite{Kurita_PRB} (triangles, top axis), deduced from the resistivity measurements up to 3.2\,GPa under zero field.
The pressure is scaled with the critical value $P_\mathrm{c}$.
PM, AFM, and SC indicate the paramagnetic, antiferromagnetic, and superconducting states, respectively. 
For the SC phase, open and solid symbols indicate $T_\mathrm{c}^\mathrm{on}$ (onset) and 
$T_\mathrm{c}^{\rho=0}$ (zero resistivity), respectively. 
$T_\mathrm{c}^{B=\mathrm{15T}}$ determined at $B$\,=\,15\,T is also shown.
Dashed curves are a guide to the eyes.
} \label{fig1}
\end{figure}

Pressure tuning of the electronic structures in stoichiometric compounds is a better means of studying a QCP 
than tuning by chemical substitution, which might obscure a QCP by the inevitably introduced randomness.
We therefore study EuFe$_2$As$_2$ under applied pressure (Fig.\,~\ref{fig1}).
The transition temperature $T_0$ is about 190\,K at ambient pressure, and the critical pressure $P_\mathrm{c}$, 
where indications of the transition at $T_\mathrm{0}$ disappear and bulk superconductivity appears, 
is 2.5\,-\,2.7\,GPa.\cite{Miclea_PRB,Terashima_Eu1,Kurita_JPCS,Kurita_PRB,Matsubayashi_PRB}
This sudden disappearance of $T_0$ is incompatible with a QCP.
Although the Eu$^{2+}$ moments exhibit an AF order at $T_\mathrm{N}$\,$\sim$\,20\,K,
the ferromagnetic (FM) alignment can be achieved at only a few Teslas\,\cite{Jiang_NJP09,Xiao_PRB2009,Xiao_PRB2010}
and the spin disorder scattering can be minimized.\cite{Terashima_Eu2}
Because of the large exchange field from the Eu$^{2+}$ moments to the conduction electron spins, 
the upper critical field $B_\mathrm{c2}$ for the $P$-induced superconductivity is much smaller\,\cite{Terashima_Eu1,Kurita_PRBR}
than that for other iron-based SCs with similar $T_\mathrm{c}$.\cite{Hunte_LaFeAsO,Yuan_BaK122,Kotegawa_Sr}
These unique characteristics of EuFe$_2$As$_2$, thus, provide a significant opportunity to experimentally investigate the iron-based superconductivity with high $T_\mathrm{c}$ of 30\,K.
Our measurements of transport properties and upper critical fields up to 3.2\,GPa reported below show 
no evidence of diverging quasiparticle effective mass at $P_\mathrm{c}$, indicating that the emergence of 
$P$-induced superconductivity in this clean system does not involve a QCP.
However, it does not curtail the importance of spin fluctuations: we observe $T$-linear resistivity 
at pressures near $P_\mathrm{c}$ and find that the Fermi liquid $T^2$ dependence 
can be recovered by the application of a magnetic field.

\section{Experimental Details}

Single crystals of EuFe$_2$As$_{2}$ were grown by the Bridgman method 
from a stoichiometric mixture of the constituent elements. 
Resistivity and Hall effect were measured simultaneously
by a conventional six-contact method with an ac current $I$ for $I$\,$\parallel$\,$ab$ and $B$\,$\parallel$\,$c$.
For the magnetotransport measurements, we used thin platelike samples ($\sim$\,1\,$\times$\,0.4\,$\times$\,0.03\,mm$^3$) 
with residual resistivity ratio $RRR$\,=\,13 ($P_\mathrm{c}$\,=2.7\,GPa\,\cite{Kurita_JPCS}).
On the other hand, samples with $RRR$\,=\,7 ($P_\mathrm{c}$\,=2.5\,GPa\,\cite{Kurita_PRB}) were 
used to analyze the $P$ dependence of $B_\mathrm{c2}$. 
As shown in Fig.\,~\ref{fig1}, 
The $P$ evolutions of $T_\mathrm{0}$, $T_\mathrm{N}$, and $T_\mathrm{c}$ 
for the two samples with different quality can be scaled by their $P_\mathrm{c}$ values.
High pressure experiments up to 3.2\,GPa were performed down to 1.6\,K in a $^4$He cryostat 
equipped with a 17\,T SC magnet using a clamped piston cylinder pressure device.\cite{PistonCell}
Daphne 7474 (Idemitsu Kosan), which remains liquid up to 3.7\,GPa at room temperature,\,\cite{Daphne7474} 
was used as a pressure-transmitting medium.
The applied pressure was determined at 4.2\,K from the change in resistance of a calibrated Manganin wire.\cite{Terashima_Eu1}
As in our previous works,\cite{Terashima_Eu1,Kurita_PRBR} $B$ denotes the externally applied field, and the magnetization 
within a sample (up to $\sim$\,0.9\,T\,\cite{Terashima_Eu2}) is neglected.

\section{Results and Discussions}

\begin{figure}
\begin{center}
\includegraphics[width=0.9\linewidth]{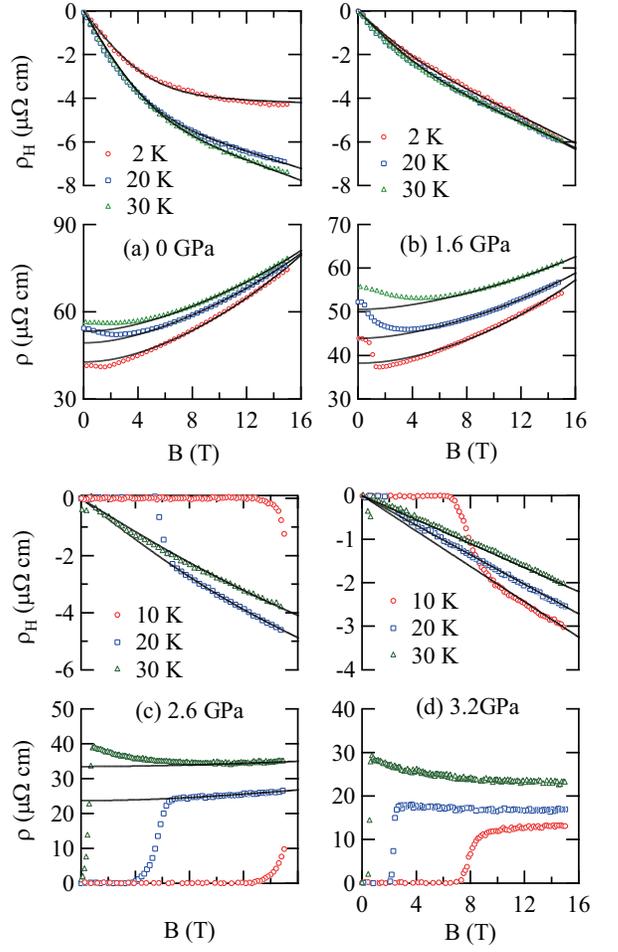}
\end{center}
\caption{(Color online) Low-temperature data of $\rho$($B$) and $\rho_\mathrm{H}$($B$) in 
EuFe$_2$As$_{2}$ ($RRR$\,=\,13) at $P$\,= (a) 0\,GPa, (b) 1.6\,GPa, (c) 2.6\,GPa, and (d) 3.2\,GPa.
Solid curves are fits to a multicarrier model.
} \label{fig2}
\end{figure}

First, we discuss how electron and hole carriers evolve as a function of $P$ via multicarrier analysis. 
Figure~\ref{fig2}(a) shows the low-$T$ data of transverse magnetoresistivity $\rho$($B$) 
and Hall resistivity $\rho_\mathrm{H}$($B$) in EuFe$_2$As$_{2}$ ($RRR$=13) at ambient pressure.
The $\rho$($B$) curves show a minimum (e.g., $\sim$\,2\,T at 2\,K), which is attributable to 
the $B$-induced FM alignment of the Eu$^{2+}$ moments.\cite{Jiang_NJP09,Xiao_PRB2009,Xiao_PRB2010}
At high fields, $\rho$($B$) shows positive magnetoresistance (MR), 
as expected from the cyclotron motion of electrons. 
$\rho_\mathrm{H}$($B$) exhibits pronounced nonlinear behavior at low temperatures.\cite{Terashima_Eu2}
The field-induced transition of the Eu$^{2+}$ moments is not detectable in the $\rho_\mathrm{H}$($B$) curves, 
indicating negligible effect of the Eu$^{2+}$ moments on the number of carriers.
At pressures sufficiently below $P_\mathrm{c}$, the $\rho_\mathrm{H}$($B$) and $\rho$($B$) curves are qualitatively similar to 
those at ambient pressure except that the curvature in $\rho_\mathrm{H}$($B$) 
and the magnitude of MR in $\rho_\mathrm{H}$($B$) decreases with increasing $P$ [Fig.~\ref{fig2}(b)].
In the vicinity of $P_\mathrm{c}$, SC transitions due to the partial [Fig.~\ref{fig2}(c)] or bulk superconductivity appears.
In the high field normal state, $\rho$($B$) and $\rho_\mathrm{H}$($B$) still slightly 
exhibit a positive MR and nonlinear behavior, respectively.
As $P$ is increased to above $P_\mathrm{c}$, $\rho_\mathrm{H}$ exhibits nearly $B$-linear dependence [Fig.~\ref{fig2}(d)],
whereas $\rho$ indicates negative MR due to the suppression of spin fluctuations of 
the Fe ions,\cite{Terashima_Eu2} except in the low-$T$ and high-$B$ regions where the cyclotron motion dominates.

\begin{figure}
\begin{center}
\includegraphics[width=0.89\linewidth]{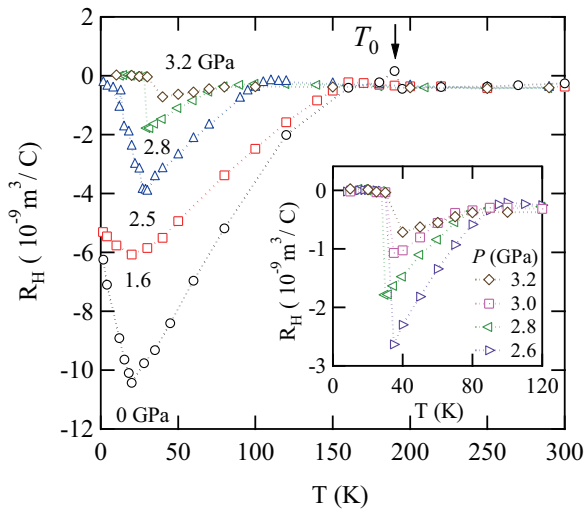}
\end{center}
\caption{(Color online) 
 $T$ dependence of the Hall coefficient $R_\mathrm{H}$ (=$d\rho_{xy}$/$dB|_{B\rightarrow 0}$) at several pressures.
The inset shows an expanded view of $R_\mathrm{H}$($T$) in the low-$T$ region at high pressures.
} \label{fig3}
\end{figure}

Figure~\ref{fig3} shows the $T$ dependence of the Hall coefficient $R_\mathrm{H}$, 
as defined by d$\rho_\mathrm{H}$/d$B$ at $B$\,=\,0, under several pressures.\cite{RH_differentB} 
The enhancement of $|R_\mathrm{H}(T)|$ 
below $T_\mathrm{0}$ for $P$\,$<$\,$P_\mathrm{c}$ indicates the destruction of substantial parts of the FS.\cite{Terashima_Eu2}
For $P$\,$>$\,$P_\mathrm{c}$ (inset), $|R_\mathrm{H}(T)|$ still increases below $\sim$\,80\,K.
Similar enhancement of $|R_\mathrm{H}(T)|$ has been observed in the paramagnetic phase of BaFe$_2$(As,P)$_2$, and 
it has been argued that the behavior cannot be explained by a multiband picture for a Fermi liquid.\cite{Kasahara_PRB2010}
However, in the present case, $|R_\mathrm{H}|$ at 2.8\,GPa is 2\,$\times$\,10$^{-9}$\,m$^3$/C ($T$\,$\sim$\,$T_\mathrm{c}$), 
which corresponds to $\approx$\,0.16\,electron/Fe (e/Fe) in a single-carrier model.
This value is comparable to the band-calculation value for BaFe$_2$As$_{2}$ (0.15\,e/Fe), 
and hence can be accounted for within a simple two-carrier picture.

\begin{figure}
\begin{center}
\includegraphics[width=0.9\linewidth]{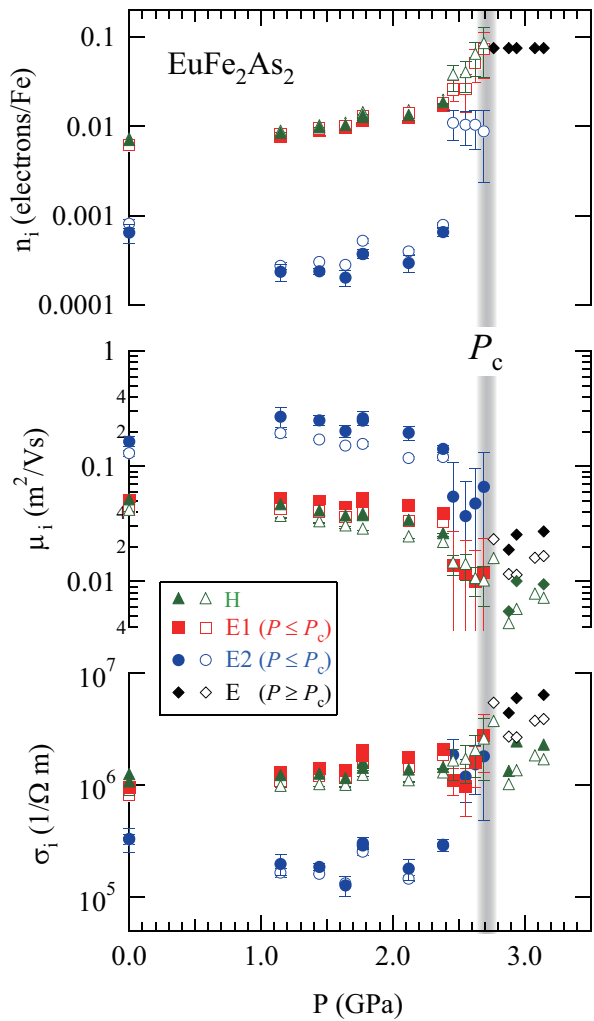}
\end{center}
\caption{(Color online) 
Pressure evolution of the carrier density $n_i$, mobility $\mu_i$, and conductivity $\sigma_i$,
deduced from multicarrier analysis.
Solid and open symbols correspond to the results obtained 
at $\le$\,10\,K and 20\,K, respectively.
One hole (H)  and two types of electrons (E1 and E2) are considered for $P$\,$\le$\,$P_\mathrm{c}$ (=\,2.7\,GPa), 
whereas a simple two-band model with 
$n_\mathrm{H}$\,=\,$n_\mathrm{E}$\,=\,7.5\,$\times$\,10$^{-2}$ electrons/Fe (e/Fe) 
is assumed for $P$\,$>$\,$P_\mathrm{c}$.
} \label{fig4}
\end{figure}

EuFe$_2$As$_{2}$ is a compensated metal with an equal number of electrons and holes, 
for which a simple two-carrier model predicts a linear $\rho_\mathrm{H}$.
The nonlinearity in $\rho_\mathrm{H}$($B$) below $P_\mathrm{c}$ thus indicates 
that more than two carriers contribute to the electronic transport.
In the case of the sister compound BaFe$_2$As$_{2}$, a Shubnikov-de Haas oscillation study has shown 
that the Fermi surface in the AF phase consists of one hole and two electron pockets.\cite{Terashima_BaFe2As2}
In keeping with this, a three-carrier model can account for the nonlinear behavior of $\rho_\mathrm{H}$($B$) 
as well as the $\rho$($B$) data for BaFe$_2$As$_{2}$.\cite{Ishida_multicarrier}

We therefore apply essentially the same three-carrier analysis to the obtained $\rho_\mathrm{H}$($B$) and $\rho$($B$) data 
for EuFe$_2$As$_{2}$ at $P$\,$\le$\,$P_\mathrm{c}$,
assuming one hole (H) and two electrons (E1 and E2) with density $n_i$ and mobility $\mu_i$.
We impose a constraint $n_\mathrm{H}=n_\mathrm{E1}+n_\mathrm{E2}$,
owing to the carrier compensation, which is held under applied pressures.
To avoid the effect of the Eu$^{2+}$ moments or superconductivity, high-field data are used for the analysis.
The solid curves in Fig.\,~\ref{fig2}(a) indicate the fits, which capture the overall features of the experimental results.
The fitting gives ($n_\mathrm{H}$, $n_\mathrm{E1}$, $n_\mathrm{E2}$; $\mu_\mathrm{H}$, $\mu_\mathrm{E1}$, $\mu_\mathrm{E2}$) 
= (6.6, 5.6, 0.92 [10$^{-2}$\,e/Fe]; 0.57, 0.57, 1.5 [10$^{3}$\,cm$^2$/V$\cdot$s]) for $T$\,=\,2\,K.
The fitting errors are approximately 3\,\% for H and E1, and 20\,\% for E2.
The parameter sets obtained at 20 and 30\,K are comparable to those at 2\,K.
One can find the tendency that $n_\mathrm{H}$\,$\approx$\,$n_\mathrm{E1}$\,$\gg$\,$n_\mathrm{E2}$ 
and $\mu_\mathrm{H}$\,$\approx$\,$\mu_\mathrm{E1}$\,$\ll$\,$\mu_\mathrm{E2}$,
similar to the case of BaFe$_2$As$_{2}$.\cite{Ishida_multicarrier}
The magnitudes of the parameter sets, particularly $\mu_i$, for EuFe$_2$As$_{2}$ ($RRR$\,=\,13)
are comparable to those for as-grown samples of BaFe$_2$As$_{2}$.\cite{Ishida_multicarrier}
For other pressures, a similar quality of fitting can be obtained.

\begin{figure}
\begin{center}
\includegraphics[width=0.9\linewidth]{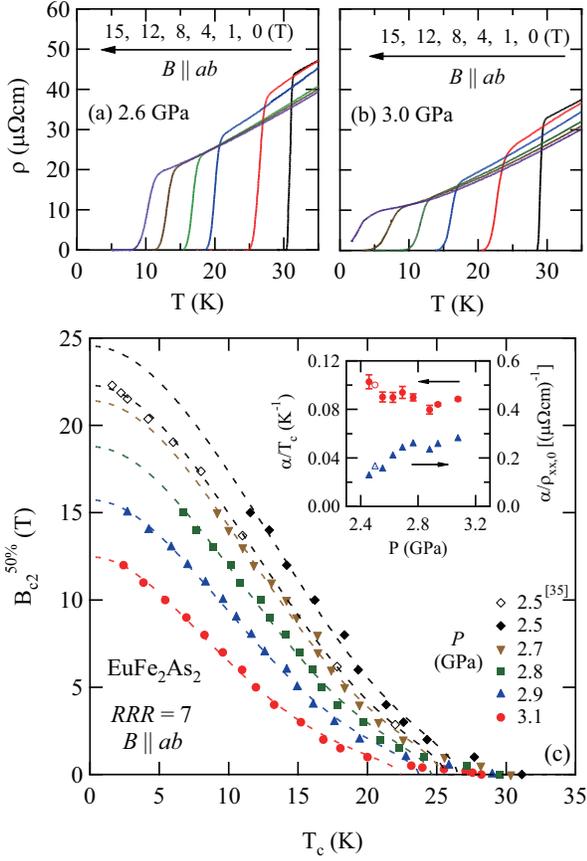}
\end{center}
\caption{(Color online) 
$\rho$ vs $T$ of EuFe$_2$As$_{2}$ ($RRR$\,=\,7, $P_\mathrm{c}$\,=\,2.5\,GPa),
under several $B$ for $B$\,$\parallel$\,$ab$ at $P$\,= (a) 2.6\,GPa and (b) 3.0\,GPa.
(c) $P$ dependence of $B_\mathrm{c2}^\mathrm{50\%}$ vs $T_\mathrm{c}$ 
of EuFe$_2$As$_{2}$ for $B$\,$\parallel$\,$ab$. 
Dashed curves are fits to the multiple pair-breaking formula (see text). 
Open symbols indicate the published data obtained from high-field resistivity measurements
at 2.5\,GPa up to 27\,T.\cite{Kurita_PRBR}
The inset shows $\alpha/T_\mathrm{c}$ and $\alpha/\rho_0$ as functions of $P$.
} \label{fig5}
\end{figure}

At $P$\,$>$\,$P_\mathrm{c}$, we assume one electron carrier (E) and one hole carrier (H)
with $n_\mathrm{H}$\,=\,$n_\mathrm{E}$.
In the analysis, the slope of $\rho_\mathrm{H}$($B$) and the $\rho$ value at $B$\,=\,0 are used.
To determine the parameter sets ($n_\mathrm{H}$,$n_\mathrm{E}$; $\mu_\mathrm{H}$, $\mu_\mathrm{E}$)  uniquely,
we fix $n_\mathrm{H}$\,=\,$n_\mathrm{E}$\,=\,7.5\,$\times$\,10$^{-2}$ e/Fe.
A band-structure calculation suggests that the carrier density for BaFe$_2$As$_{2}$ is 0.15\,e/Fe.\cite{Fang_PRB2009}
However, the shrinking of the FS has been observed in BaFe$_2$(As, P)$_2$\,\cite{Shishido_PRL2010} 
and has been theoretically attributed to strong interband scattering.\cite{Ortenzi_PRL2009}
The volume of the FS is approximately halved as 
the optimal doping is approached in BaFe$_2$(As,P)$_2$.\cite{Shishido_PRL2010}
We therefore use the halved value.

Figure~\ref{fig4} displays the $P$ evolutions of $n_i$, $|\mu_i|$, and conductivity $\sigma_i$. 
As the pressure approaches $P_\mathrm{c}$, $n_i$ increases, while $|\mu_i|$ decreases.
It appears that $n_i$ and $|\mu_i|$ develop reasonably continuously to their values at $P$\,$>$\,$P_\mathrm{c}$.
Neither of the $P$ dependences of $\mu_i$ (= $e\tau_i/m_i^*$) or $\sigma$ (= $e^2 \tau_i n_i/m_i^*$) suggests 
the divergence of $m^*$ or $(n/m^*)^{-1}$ at $P_\mathrm{c}$.

\begin{figure}
\begin{center}
\includegraphics[width=0.9\linewidth]{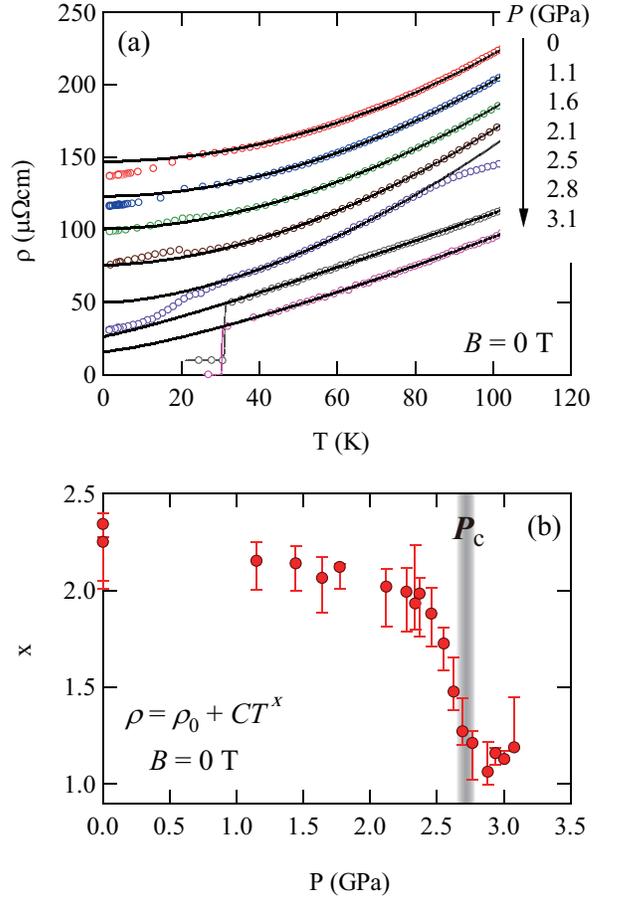}
\end{center}
\caption{(Color online) 
(a) $\rho$ vs $T$ under zero field at several pressures.
The data are arbitrarily shifted in the longitudinal direction for clarity.
Solid curves are fits to $\rho$\,=\,$\rho_{0}$\,+\,$CT^{x}$.
(b) $P$ evolution of the exponent $x$ under zero field. 
Solid symbols are obtained from fits in the $T$ range between 35 and 60\,K.
Error bars are estimated from fits in several $T$ ranges from 35\,K ($>$\,$T_\mathrm{N}, T_\mathrm{c}$) 
to temperatures between 50\,K and $T_\mathrm{0}$.
} \label{fig6}
\end{figure}

We further substantiate the absence of a QCP by deriving the $P$ dependence of the effective masses 
from $B_\mathrm{c2}$-$T_\mathrm{c}$ phase diagrams under applied pressures.
Figures~\ref{fig5}(a) and (b) show $\rho$ vs $T$ at 2.6 and 3.0\,GPa, respectively, 
in the $RRR$\,=\,7 sample under several $B$ for $B$\,$\parallel$\,$ab$.
$T_\mathrm{c}$ under each magnetic field is determined by the midpoint temperature of the SC transitions.
Figure~\ref{fig5}(c) shows the thus determined upper critical field $B_\mathrm{c2}$  
as a function of temperature for several pressures.
It appears that $B_\mathrm{c2}(0)$ is highest at $P$\,$\sim$\,$P_\mathrm{c}$ and decreases with increasing $P$.
In EuFe$_2$As$_{2}$, orbital and Pauli paramagnetic effects and magnetic Eu$^{2+}$ moments 
all play an important role in determining $B_\mathrm{c2}$, which complicates the understanding of 
the obtained $B_\mathrm{c2}$ vs $T_\mathrm{c}$.
In a previous paper, we analyzed $B_\mathrm{c2}(T)$ data obtained at 2.5\,GPa (also shown in Fig.~\ref{fig5}(c)) 
using a multiple pair-breaking model\,\cite{WHH1966,JP_effect2}
that includes the antiferromagnetic exchange field $B_J$ due to magnetic Eu$^{2+}$ moments.
We obtained the spin-orbit scattering parameter $\lambda_\mathrm{so}=2.4$, 
and the maximum of $|B_J|$ as $B_J^\mathrm{m}=75$\,T,
where the Maki parameter $\alpha$\,=\,3 is fixed.\cite{Kurita_PRBR}
It is known that $m^*$ is related to $\alpha$ through
$m^* \propto \sqrt{ \alpha/T_\mathrm{c}}$ or $\propto \alpha/\rho_0$ 
($\rho_0$: residual resistivity) in the clean or dirty limit, respectively.
Thus, we estimate $\alpha$ as a function of $P$ using the same model.
As $\alpha$ is highly sensitive to other fitting parameters,
we use the values of $\lambda_\mathrm{so}$\,=\,2.4 and $B_J^\mathrm{m}$\,=\,75\,T for all the pressures.
Dashed curves are fitting results using $\alpha$ and $T_\mathrm{c}$ as free parameters.
The inset of Fig.~\ref{fig5}(c) shows the $P$ dependences of $\alpha/T_\mathrm{c}$ and $\alpha/\rho_0$.
The former exhibits only a modest increase, whereas the latter exhibits a decrease
as $P_\mathrm{c}$ is approached.
This indicates that, either in the clean or dirty limit,
there is no divergence of $m^*$ as the pressure approaches $P_\mathrm{c}$.

\begin{figure}
\begin{center}
\includegraphics[width=0.9\linewidth]{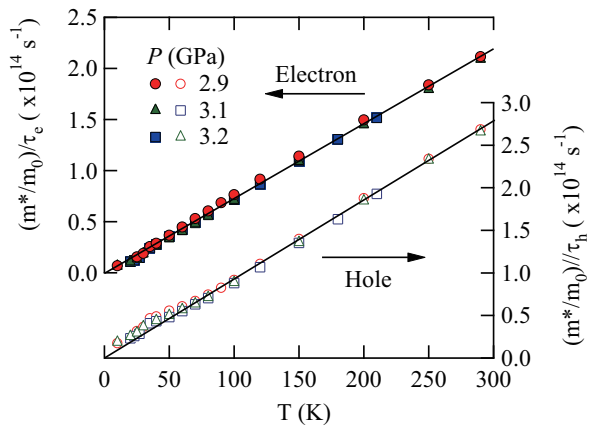}
\end{center}
\caption{(Color online) 
$(m^*/m_0)/\tau_i$ vs $T$ for electron ($i$=E) and hole ($i$=H)
at 2.9, 3.1, and 3.2\,GPa ($>P_\mathrm{c}$).
Solid lines are a guide to the eyes.
} \label{fig7}
\end{figure}

We now address the issue of the NFL behavior.
Figure~\ref{fig6}(a) shows the $T$ dependence of $\rho$ under zero field 
in EuFe$_2$As$_{2}$ ($RRR$\,=\,13) at several pressures.
Solid curves are fits to $\rho$\,=\,$\rho_{0}$\,+\,$CT^{x}$ ($C$: constant).
To avoid the effect of Eu$^{2+}$ moments or superconductivity,
we use data in a fitting range from 35\,K (fixed) to several temperatures between 50\,K and $T_0$.
Figure~\ref{fig6}(b) indicates the $P$ evolution of the exponent $x$.
At $P$\,$\ll$\,$P_\mathrm{c}$, 
Fermi liquid (FL) like behavior ($x$\,$\sim$\,2) is observed.
As the pressure approaches $P_\mathrm{c}$, $x$ decreases rapidly and reaches approximately unity.

It has previously been proposed that the $T$-linear resistivity in iron-based SCs arises from the
$T$-dependent carrier concentration and that the carrier scattering rate $\tau^{-1}$ obeys 
a standard FL $T^2$ law.\cite{Fang_PRB2009,Albenque_PRL}
We therefore show the $T$ dependence of $(m^*/m_0)/\tau_i=(e/m_0)\mu_i^{-1}$
at 2.9, 3.1, and 3.2\,GPa ($>P_\mathrm{c}$) obtained from the above two-carrier analyses in Fig.~\ref{fig7}.
The figures indicate that the scattering rates for both electrons and holes are 
nearly proportional to $T$ in a remarkably wide $T$ range.
Although we have assumed $n_\mathrm{H}$\,=\,$n_\mathrm{E}$\,=\,7.5\,$\times$\,10$^{-2}$ e/Fe 
in estimating $\mu_i$ as noted above, any carrier number in the range between 0.05 and 0.1 e/Fe 
gives a similar approximately $T$-linear dependence of $\mu_i$.  

\begin{figure}
\begin{center}
\includegraphics[width=0.9\linewidth]{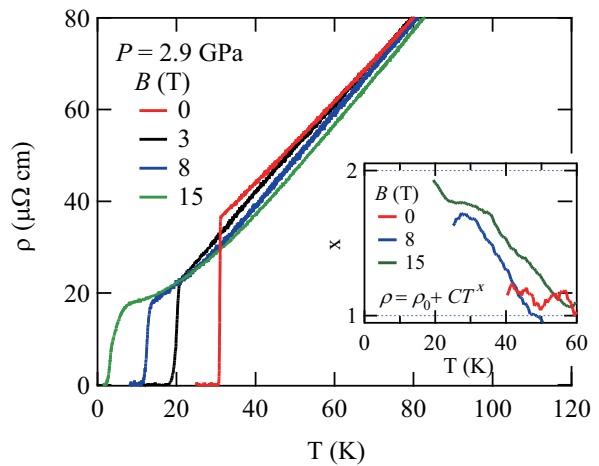}
\end{center}
\caption{(Color online) 
$\rho$($T$) at 2.9\,GPa under 0, 3, 8, and 15\,T.
The inset shows the $T$ dependence of $x$ under several fields.
} \label{fig8}
\end{figure}

Figure~\ref{fig8} shows $\rho(T)$ at 2.9\,GPa ($\sim$\,$P_\mathrm{c}$) under several fields of up to 15\,T.
The FL $T^2$ behavior is gradually restored with increasing $B$.
To estimate the $T$ variation of the exponent $x$ under applied fields,
we fit the data to $\rho$\,=\,$\rho_{0}$\,+\,$CT^{x}$ in several temperature ranges.
The inset of Fig.~\ref{fig8} shows the $T$ variation of $x$ under 0, 8, and 15\,T.
Under zero field, the $x$ value is close to one.
Under applied fields of 8 and 15\,T, with decreasing $T$, 
the $x$ value increases from $\sim$\,1 and approaches $\sim$\,2.
This clearly indicates that the origin of the $T$-linear resistivity is spin fluctuations.
As is well known, spin fluctuation theories predict $T$-linear resistivity 
for two-dimensional nearly AF metals.\cite{Moriya_JPSJ1990}
To our knowledge, there has been no observation of a $B$-induced change in the resistivity exponent from 1 to 2 in iron-based SCs.
The present observation is most likely to result from the fact that 
the conduction carrier spins are influenced by the large exchange field from the Eu$^{2+}$ moments, 
in addition to the externally applied field.
That is, as soon as the Eu$^{2+}$ moments are fully aligned by the applied field, 
the conduction electrons feel a large exchange field of $-B_J^\mathrm{m}$\,=\,$-$75\,T, 
which effectively suppresses the spin fluctuations. 

\section{Conclusions}

To conclude, our analyses of magnetotransport and upper critical fields in EuFe$_2$As$_{2}$ 
under high pressure indicate that there is no QCP at $P_\mathrm{c}$ in this pure compound, 
which is in sharp contrast to the observation in BaFe$_2$(As,P)$_2$.\cite{Hashimoto_Science2012,Walmsley_PRL2013}
On the other hand, we have shown that the scattering rates for both electrons and holes are 
approximately $T$-linear for $P$\,$>$\,$P_\mathrm{c}$.
The recovery of the FL $T^2$ dependence of $\rho$ at high fields clearly indicates that 
spin fluctuations are the origin of the anomalous scattering.
It appears that systematic analyses of spin (and/or orbital) fluctuations based on the electronic structures 
of individual materials, beyond a generic scenario based on a QCP, are necessary to elucidate 
the mechanism of iron-based superconductivity.

\section*{Acknowledgment}
We would like to thank H. Eisaki, S. Ishida, and H. Harima for fruitful discussions.
This work was partially supported by a Grant-in-Aid for Young Scientists (No. 23740279) 
from the Ministry of Education, Culture, Sports, Science and Technology, Japan.

\section*{Appendix}

The procedure for the three carrier analysis used in simultaneously fitting 
the obtained $\rho$ and $\rho_\mathrm{H}$ data
of EuFe$_2$As$_2$ is shown below. 
This procedure is essentially the same as that used in the recent work on 
the sister compound BaFe$_2$As$_2$.\cite{Ishida_multicarrier}

Tensor components of the electrical conductivity, $\sigma_{xx}$ (=$\sigma_{yy}$) and $\sigma_{xy}$ (=$-\sigma_{yx}$), for three carriers can be expressed in the following forms using those of the electrical resistivity, $\rho_{xx}$ (=$\rho_{yy}$=$\rho$) and $\rho_{xy}$ (=$-\rho_{yx}$=$\rho_\mathrm{H}$):

\begin{equation}
\begin{aligned}
{\sigma}_{xx}=\label{eq1}\frac{{\rho}_{xx}}{{\rho}_{xx}^2+{\rho}_{xy}^2}=\sum^3_{i=1}\frac{q_in_i{\mu}_i}{1+({\mu}_iB)^2}
\\
{\sigma}_{xy}=\frac{{\rho}_{xy}}{{\rho}_{xx}^2+\rho_{xy}^2}=\sum^3_{i=1}\frac{q_in_i{\mu}_i^2B}{1+({\mu}_iB)^2}
\end{aligned}
\end{equation}

\noindent
where $q_i$, $n_i$, and $\mu_i$ are the charge, density, and mobility of the $i$th carrier, respectively, 
and the tensors of the electrical conductivity $\bm{\sigma}$ and resistivity $\bm{\rho}$ have the forms:\\

\begin{equation}
\bm{\sigma}=
\begin{pmatrix}
{\sigma}_{xx} & {\sigma}_{xy} \\
{-\sigma}_{xy} & {\sigma}_{xx}
\end{pmatrix}
,\hspace{5mm}
\bm{\rho}=
\begin{pmatrix}
{\rho}_{xx} & {\rho}_{xy} \\
{-\rho}_{xy} & {\rho}_{xx}
\end{pmatrix}
\end{equation}

From Eq. (1), $\rho_{xx}$ and $\rho_{xy}$ can be written as follows:

\begin{equation}
\begin{aligned}
{\rho}_{xx}=\frac{\sum^3_{i=1}\frac{q_in_i{\mu}_i}{1+({\mu}_iB)^2}}{\left[\sum^3_{i=1}\frac{q_in_i{\mu}_i}{1+({\mu}_iB)^2}+\sum^3_{i=1}\frac{q_in_i{\mu}_i^2B}{1+({\mu}_iB)^2}\right]}
\\
{\rho}_{xy}=\frac{\sum^3_{i=1}\frac{q_in_i{\mu}_i^2B}{1+({\mu}_iB)^2}}{\left[\sum^3_{i=1}\frac{q_in_i{\mu}_i}{1+({\mu}_iB)^2}+\sum^3_{i=1}\frac{q_in_i{\mu}_i^2B}{1+({\mu}_iB)^2}\right]}\\
\end{aligned}
\end{equation}
\\

\end{document}